# Real-space imaging of acoustic plasmons in large-area CVD graphene

Sergey G. Menabde,[1,†] In-Ho Lee,[2,†] Sanghyub Lee,[3,4] Heonhak Ha,[1] Jacob T. Heiden,[1] Daehan Yoo,[2] Teun-Teun Kim,[3,5] Young Hee Lee,[3,4] Tony Low,[2] Sang-Hyun Oh,[2,*] and Min Seok Jang[1,**]

[1]School of Electrical Engineering, Korea Advanced Institute of Science and Technology (KAIST), Daejeon, Korea
[2]Department of Electrical and Computer Engineering, University of Minnesota, Minneapolis, USA
[3]Center for Integrated Nanostructure Physics (CINAP), Institute for Basic Science (IBS), Suwon, Korea
[4]Department of Energy Science, Sungkyunkwan University, Suwon, Korea
[5]Department of Physics, University of Ulsan, Ulsan, Korea
[*]sang@umn.edu
[**]jang.minseok@kaist.ac.kr
[†]These authors contributed equally to this work

**ABSTRACT**

An acoustic plasmonic mode in a graphene-dielectric-metal heterostructure has recently been spotlighted as a superior platform for strong light-matter interaction. It originates from the coupling of graphene plasmon with its mirror image and exhibits the largest field confinement in the limit of a nm-thick dielectric. Although recently detected in the far-field regime, optical near-fields of this mode are yet to be observed and characterized. Direct optical probing of the plasmonic fields reflected by the edges of graphene via near-field scattering microscope reveals a relatively small damping rate of the mid-IR acoustic plasmons in our devices, which allows for their real-space mapping even with unprotected, chemically grown, large-area graphene at ambient conditions. We show an acoustic mode that is twice as confined – yet 1.4 times less damped – compared to the graphene surface plasmon under similar conditions. We also image the resonant acoustic Bloch state in a 1D array of gold nanoribbons responsible for the high efficiency of the far-field coupling. Our results highlight the importance of acoustic plasmons as an exceptionally promising platform for large-area graphene-based optoelectronic devices operating in mid-IR.

**INTRODUCTION**

An acoustic plasmonic mode supported by a system of two graphene sheets[1], or by a single graphene sheet over a metal gate[2], was experimentally detected only recently[3], but has already garnered significant attention due to its unprecedented field confinement in the highly sought-after mid-infrared (mid-IR) regime[4-7]. This mode is supported by a heterostructure comprising of a metal, a dielectric spacer, and a graphene layer, where the image charges in the metal effectively "mirror" the charge density oscillations in the doped graphene layer. The acoustic graphene plasmon (AGP) supported by the heterostructure is mostly confined in the dielectric spacer and does not experience a cutoff as the spacer thickness decreases[5], thus resembling the fundamental plasmonic mode of a narrow metal gap. The AGP excited at the important mid-IR frequencies[8,9] does not exhibit significant loss and is detectable even when the spacer is reduced to a single atomic layer of hexagonal boron nitride[7]. Inside such a narrow dielectric spacer, the AGP wavevector is about two orders of magnitude larger than that of free space, which grants access to quantum and non-local phenomena in graphene[7,10,11], and allows for the AGP localization in nanostructures[12] with a stunning mode volume confinement factor[4] of ~$10^{10}$. This ultimate capability to compress mid-IR light significantly outperforms that of all other polaritonic species in van der Waals materials[8], including graphene surface plasmon[13,14] (GSP), and is crucial for applications that require strong light-matter interaction such as molecular sensing[6,15-18], polaritonic dispersion engineering in van der Waals crystals[19,20], and dynamic light manipulation by graphene-based active metasurfaces[21-24].

The key advantage of the AGP is its confinement within the (low-loss) dielectric spacer, in contrast to the GSP bound to the graphene layer. Therefore, ohmic losses in graphene are expected to hinder the AGP propagation to a lesser extent compared to GSP. On the other hand, the larger AGP wavevector requires an intermediary structure to alleviate the phase mismatch under the far-field excitation. So far, heterostructures containing an array of metallic elements have been used to couple far-field radiation to the AGP mode[4,6,7]. However, the near-field optical probing of AGP is yet to be demonstrated.

In this work, we employ a scattering-type scanning near-field optical microscope (s-SNOM) based on an atomic force microscope (AFM) for real-space mapping and analysis of mid-IR AGP. We directly measure the AGP dispersion, evaluate the mode propagation loss, and investigate its behavior in a periodic structure designed for the far-field coupling. Most importantly, our results reveal a small damping rate of infrared AGP even when unprotected, chemically grown, large-area graphene is used at ambient conditions, suggesting an exceptionally promising prospects for large-area graphene-based optoelectronic devices.

A periodic array of gold nanoribbons underneath graphene provides a resonant far-field coupling to AGP[6]. We also image the acoustic modes in such an array, including the resonant AGP Bloch state responsible for the high coupling efficiency in far-field regime. The near-field data indicates that the local density of plasmonic states does not increase at the frequency of the array resonance. Numerical analysis of the AGP Bloch state reveals that the AGP spectrum acquires a blue shift with respect to the uniform AGP case.

**RESULTS**

**Near-field coupling to acoustic graphene plasmon**

Although the AGP fields are mainly confined inside the dielectric spacer, their evanescent components have non-zero amplitudes beyond the graphene layer (Fig. 1a). Particularly, the vertical component of the mode's electric field, $E_z$, penetrates into the free-space above the structure, and hence, can couple to the AFM tip of the s-SNOM effectively acting as a z-oriented electric dipole[25-28] (Fig. 1b). We define the $E_z$ penetration depth above graphene, $D_e$, as corresponding to the 1/e attenuation of the field amplitude. By solving Maxwell's equations in a multilayer configuration[29], it is possible to find the unique solution for the AGP eigenmode supported by the structure of interest (see Supplementary Information section S-1). Then, $D_e = 1/\text{Im}\{k_z\}$, where $k_z$ is the z-component of the AGP wavevector in the medium above graphene. At a given graphene Fermi level, $E_F$, the wavevector depends on both the excitation frequency $\omega$ and the spacer thickness $t$, and so does the $D_e$. At the same time, the amplitude of the scattered near-field signal from the AFM tip is proportional to $E_z$ at the position of the tip. Therefore, the performance of the s-SNOM method is expected to vary significantly depending on the experimental conditions.

In order to estimate the optimal experimental conditions for the near-field AGP probing by s-SNOM, it is instructive to calculate $D_e(\omega,t)$ for the structure of interest. Figure 1c shows $D_e(\omega,t)$ calculated for the range of mid-IR frequencies and spacer thicknesses, assuming $E_F = 0.5$ eV. Considering the average tip height above the sample 40~80 nm, which is approximately equal to the tip tapping amplitude[27,30], our educated guess is that the most favorable experimental conditions are expected in the frequency window below 1300 cm$^{-1}$ and above 1000 cm$^{-1}$ where Al$_2$O$_3$ absorption starts to take effect, while $t > 10$ nm.

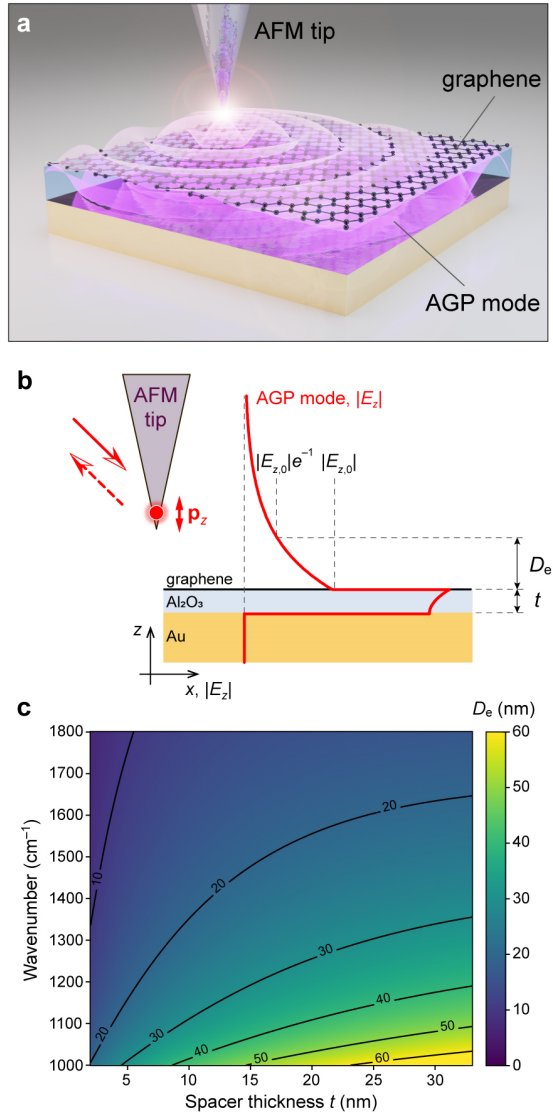

Figure 1. Plasmon coupling to the AFM tip. **a** The AFM tip couples to the AGP inside the heterostructure via the evanescent field above graphene. **b** The exponentially decaying *z*-component of the AGP electric field $E_z$ couples to the AFM tip that acts as a *z*-oriented electric dipole. **c** Penetration depth of $|E_z|$ as a function of excitation frequency and spacer thickness calculated for graphene $E_F = 0.5$ eV.

Substrates with gold and alumina films were fabricated using the template-stripping method[31,32], allowing a sub-nm roughness of all interfaces even when the gold film is patterned[6] (see Supplementary Information section S-2). The large-area monocrystalline graphene[33] was grown by chemical vapor deposition (CVD), wet-transferred on top of the Au/Al$_2$O$_3$ substrates, and chemically doped (see Methods section for details).

**Dispersion and loss analysis**

Near-field imaging of doped graphene on the Au/Al$_2$O$_3$ substrate reveals an abundance of μm-long edges of monocrystalline areas with AGP interference fringes formed due to its reflection from the graphene termination (Fig. 2a). Figure 2b demonstrates a close-up scan of such an edge at a sample with $t = 18$ nm at $\omega = 1150$ cm$^{-1}$. The near-field signal intensity $s(x,y)$ is proportional to the amplitude of the z-component of the electric field under the tip[28], $s \propto |E_z|$, thus it can be numerically calculated by full-wave simulations in a quasi-static approximation[27,34] (see Supplementary Information section S-3). As shown in Fig. 2c, the full-wave simulations by finite element method (FEM) with AFM tip modelled as a point dipole source provide a perfect fit to the interference pattern, where numerical and experimental data normalized by that from the graphene-free area. The fitted value of the optical conductivity of graphene (given by the random phase approximation in local limit[35-37]) corresponds to $E_F = 0.51$ eV and carrier mobility $\mu = 2000$ cm$^2$/Vs, consistent with a high-quality CVD graphene. The roughness-mediated scattering of AGP is neglected in the full-wave simulations. Additional simulations support this assumption, considering the measured RMS roughness of 0.5 nm at gold and alumina surfaces (Supplementary Information section S-2).

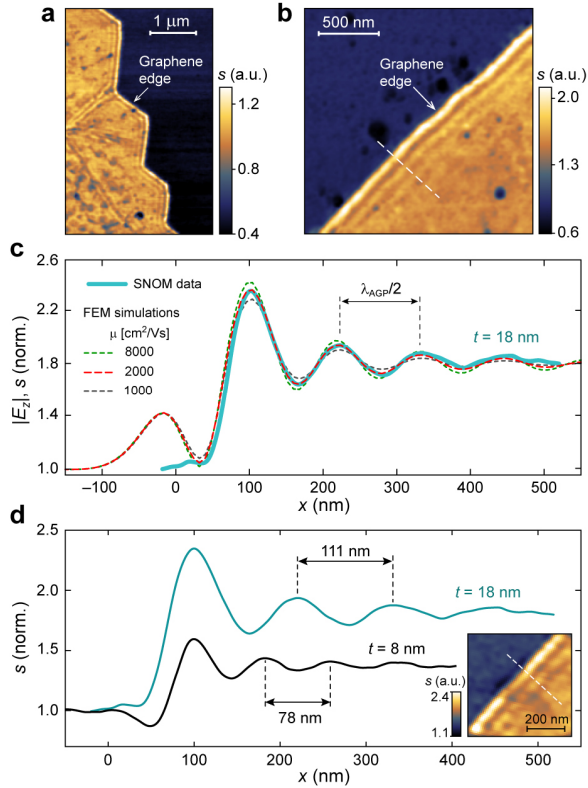

Figure 2. Near-field mapping of the AGP interference fringes at the graphene edge. **a** Distribution of the near-field signal intensity $s(x,y)$ over the doped graphene near its edge: the monocrystalline CVD graphene naturally provides long high-quality edges where the AGP interference fringes are visible. **b** High resolution $s(x,y)$ scan of the graphene edge with AGP interference fringes. **c** Near-field signal intensity (blue solid) measured across the edge shown in **b** along the white dashed line (averaged over a ten-pixels-wide line), and calculated $|E_z|$ at different carrier mobility in graphene (dashed) with $E_F = 0.51$ eV. **d** AGP fringes across the graphene edge in samples with spacer thickness $t = 18$ nm (blue) and 8 nm (black; fitted $E_F = 0.49$ eV). Inset: $s(x,y)$ over the sample with $t = 8$ nm. All data is at $\omega = 1150$ cm$^{-1}$.

The AGP interference fringes allow for the direct measurement of the plasmonic wavelength[28,38] $\lambda_{AGP} \approx 222$ nm (Fig. 2c,d); note that this is not the case for graphene patches of finite size where multiple reflections contribute to near-field signal[28]. Additionally, we observe AGP interference in the sample with $t = 8$ nm (inset in Fig. 2d). As expected, the AGP mode in the thinner spacer is more confined, hence the shorter $\lambda_{AGP} \approx 156$ nm and the weaker amplitude of the near-field signal, despite the similar doping level $E_F = 0.49$ eV. Furthermore, the values of $\lambda_{AGP}$ demonstrate that the observed interference fringes are that of the AGP and not the GSP in the graphene sheet, since the latter would require the Fermi level of 0.31 eV and 0.23 eV, inconsistent with the same chemical doping procedure for both samples.

By plugging the recovered graphene conductivity into the semi-analytic eigenmode solver, the parameters of the detected AGP can be readily obtained. For the sample with $t = 18$ nm (8 nm), the effective index of the AGP is $q_{AGP} = k_{AGP}/k_0 = 39.06 + 2.92i$ ($55.96 + 4.17i$), where $k_{AGP}$ is the AGP propagation constant, and $k_0$ is the free-space wavevector. It can be immediately noted that the damping rate $\gamma_{AGP} = \text{Im}\{q_{AGP}\}/\text{Re}\{q_{AGP}\}$ = 0.075 (in both samples) is very small compared to the reported value for mid-IR GSP in CVD graphene on $SiO_2$ substrate[39] $\gamma_{GSP} = 0.135$. Calculations for the same graphene on a thick $Al_2O_3$ substrate give $q_{GSP} = 23.41 + 2.47i$ ($24.43 + 2.62i$), and thus, an expected damping rate $\gamma_{GSP} \approx 0.106$ in both cases. Therefore, our near-field measurements indicate that, while the detected AGP is ×1.7 (×2.3) times *more compressed* in terms of the wavevector, it is at the same time 1.4 times *less damped* than the GSP in the same graphene sheet. Although these peculiar features of AGP are predicted by theory (see Supplementary Information section S-1), the experimental observation of the low-loss AGP with unprotected CVD graphene at ambient conditions is very encouraging for the development of large-area polaritonic devices operating in mid-IR[8].

We attribute the low AGP damping rate in our experiments to the absence of roughness-mediated scattering and the monocrystalline structure of graphene. Furthermore, we speculate that AGP in general is less sensitive to the major loss channels in graphene: the acoustic phonons and the charge impurities, which concentration must be especially high due to the chemical doping. Particularly, the electron scattering via impurities is expected to be less severe since the metal screens the impurities potential, as has been shown in the context of electron transport in graphene[40].

As demonstrated earlier, the AGP dispersion can be directly measured from the near-field images at different frequencies. The AGP dispersion measured in a sample with $t = 21$ nm (circles) is shown in Fig. 3a, along with the calculated dispersion for $E_F = 0.46$ eV. The near-field data is obtained from a series of measurements over the same sample area, which makes it possible to compare the spectral dependency of the near-field contrast[27,30] $\eta(x,y) = \frac{s(x,y)}{s_{\text{ref}}} e^{i(\varphi(x,y) - \varphi_{\text{ref}})}$, where $s_{\text{ref}}$ and $\varphi_{\text{ref}}$ are the amplitude and phase of the near-field signal over the graphene-free area, respectively. Figure 3b demonstrates mapping of $|\eta(x,y)| = s(x,y)/s_{\text{ref}}$ and corresponding AGP interference fringes at different frequencies, indicating the effective mode index increasing from 34 at 1080 cm$^{-1}$ to 46 at 1260 cm$^{-1}$. Furthermore, the spectral dependency of $|\eta|$ above graphene (averaged across the area far from the edge) generally follows the calculated value of $D_e$ (Fig. 3c), in agreement with the stronger mode confinement inside the spacer. According to Fig. 3c, the near-field contrast approaches unity when $D_e \approx 25$ nm. Therefore, based on the calculations for $D_e$ shown in Fig. 1c, we predict that the s-SNOM technique would be feasible for AGP probing even when the spacer thickness is reduced down to a few nanometers if $\omega$ is sufficiently low.

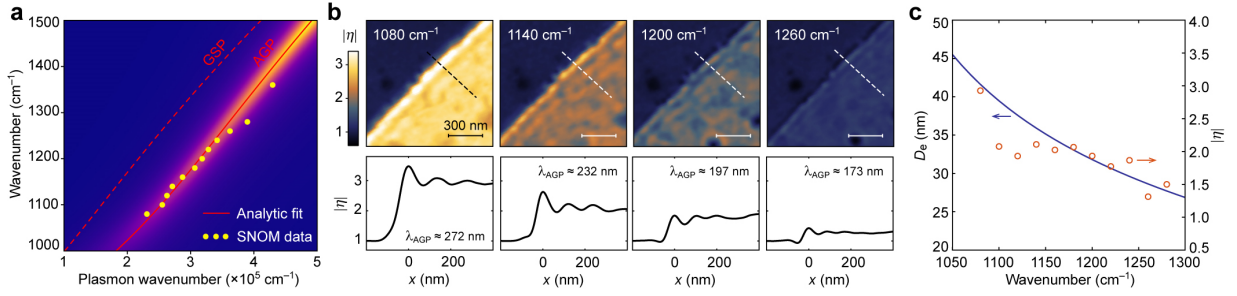

Figure 3. Plasmon dispersion. **a** AGP dispersion obtained from the interference fringes in near-field at the sample with $t$ = 21 nm (circles), and the analytically calculated dispersion for $E_F$ = 0.46 eV: the exact solutions for AGP (red solid), GSP (red dashed; for thick alumina layer), and the imaginary part of the reflection coefficient (color map). **b** Top row: distribution of the near-field contrast $|\eta(x,y)|$ over the same graphene edge obtained at different excitation frequencies. Bottom row: corresponding $|\eta|$ profiles across the graphene edge, measured along the dashed line (average value for ten-pixel-wide lines). **c** Spectral dependency of the calculated $D_e$ (solid) and the measured $|\eta|$ above the sample shown in **b** (circles; averaged over the area far from the edge).

**Acoustic graphene plasmons in periodic structure**

The uniform heterostructure is a perfect platform to probe a "pristine" AGP mode. However, due to the significant momentum mismatch, the efficient AGP coupling to a far-field requires a mediator – an array of metallic elements (e.g. gold nanoribbons), which can provide up to a 100% coupling efficiency when combined with an optical cavity[6]. We fabricated AGP resonators similar to those used in Ref.[6], where gold nanoribbons are embedded in the alumina layer (Fig. 4a). Due to the periodicity and finite width of the nanoribbons, their interaction with AGP may produce non-trivial near-field patterns depending on the ratio between the AGP wavelength, array period $P$, and ribbons width $w$. We investigate samples with different $w$, while the gap size is 30 nm and $t$ = 18 nm in all devices; the effect of the underlying cavity is not considered in this study.

The near-field signal from a non-uniform structure bears information from multiple scattering sources. Therefore, the AGP damping rate cannot be extracted from the interference fringes. At the same time, $\lambda_{AGP}$ depends solely on $E_F$ in graphene, which can be fitted using the spectral dependence of $\lambda_{AGP}$ (see Supplementary Information section S-4). Figure 4b demonstrates the spatial distribution of $|\eta|$ at different frequencies, measured over the same area of graphene deposited on alumina with embedded gold nanoribbons ($w$ = 240 nm). Dispersion fit to $\lambda_{AGP}$ at different frequencies (Fig. 4c) provides $E_F$ = 0.58 eV, while the measured effective index increases from 32 at 1080 cm$^{-1}$ to 44 at 1280 cm$^{-1}$. The correlation between $|\eta|$ and $D_e$ (Fig. 4d) is very similar to that observed in the uniform heterostructure, indicating a stronger AGP confinement at higher frequencies, while $|\eta|$ approaches unity at $D_e \approx$ 30 nm. This supports our earlier assumption that the s-SNOM method would allow for the direct optical probing of AGP even in samples with an atomically thin spacer at sufficiently low excitation frequencies.

In our experiments, the plane of incidence of the TM-polarized excitation beam is always orthogonal to the nanoribbons in order to maximize the scattering at the metal edges (as indicated by the red arrow in the first panel of Fig. 4b). While the AFM tip is able to excite the AGP with an arbitrary direction and magnitude of the wavevector, the excitation beam is expected to couple only to the mode propagating in the periodic structure across the ribbons, with maximum coupling efficiency reached at the phase-matching condition between the AGP and the array[6] $k_{AGP} = 2\pi/P$. Unexpectedly, the near-field contrast over the nanoribbons (measured far from the graphene edge; Fig. 4d) does not show any noticeable peak around the phase-matching frequency of 1105 cm$^{-1}$. To understand this and gain an insight into the near-field excitation of AGP in the array, we proceed with an analysis of the plasmonic band structure along both $y$- and $x$-direction.

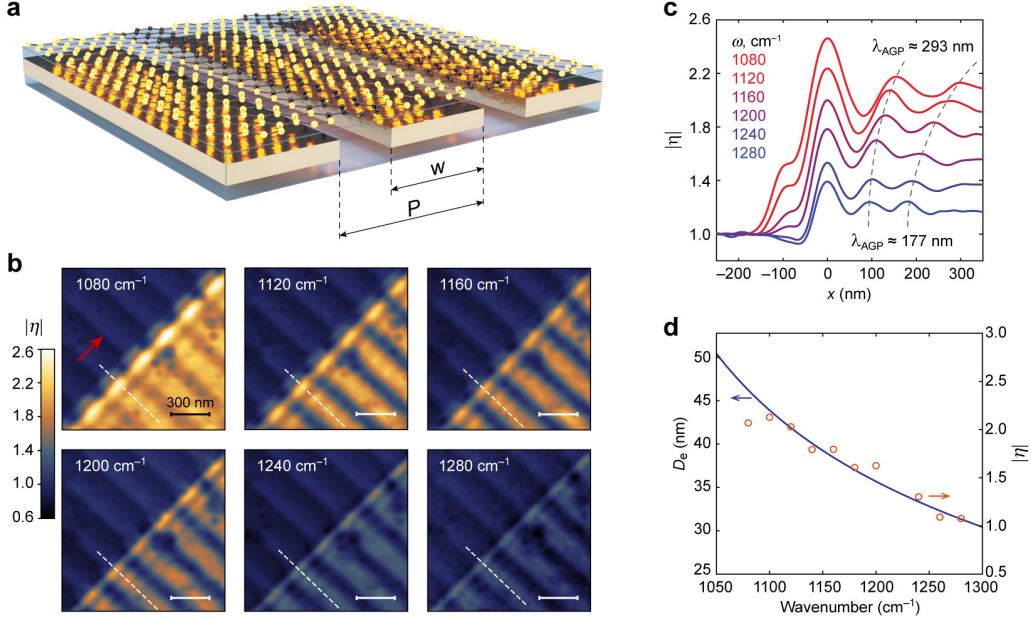

Figure 4. Near-field imaging of AGP over the periodic array of gold nanoribbons. **a** Schematics of the heterostructure with gold nanoribbons of width $w$ arranged in an array with period $P$. **b** Near-field contrast $|\eta(x,y)|$ obtained at different frequencies over the same sample area; $P = 270$ nm, $w = 240$ nm, and $t = 18$ nm. The red arrow indicates the tip illumination direction. **c** Profile of $|\eta|$ across the graphene edge along the dashed lines in **b** showing smaller AGP wavelength and $|\eta|$ at higher excitation frequencies, indicating the stronger AGP confinement. **d** Spectral dependency of maximal $|\eta|$ above graphene measured far from the edge (circles) and calculated $D_e$ (solid) for the sample shown in **b**.

The infinite 1D array of nanoribbons is modelled as a unit cell of size $P$ with periodic boundary conditions in $x$-direction (Fig. 5a). Eigenmode analysis at the frequencies of interest reveals three AGP modes propagating along the ribbons, illustrated by their field distribution in Fig. 5a-c. The "fundamental" mode (Fig. 5a) possesses the wavenumber $\beta_y^{(a)}$ equal to that of the AGP in a uniform heterostructure $\beta_{AGP}$, as demonstrated by the dispersion plot in Fig. 5d. Upon the reflection at the graphene edge, this mode forms the interference fringes, revealing the AGP wavelength at given frequency (Fig. 4b,c). The second-order array mode has two branches with different field distribution across the unit cell (Fig. 5b,c). The mode with slightly larger wavenumber $\beta_y^{(b)}$ does not couple to the nanogaps (Fig. 5b), while the other mode with $\beta_y^{(c)}$ is coupled to the metal gaps (Fig. 5c) and therefore, is more lossy. Due to their higher loss, the high-order modes are difficult to observe upon the far-field excitation, while expected to be visible in near-field.

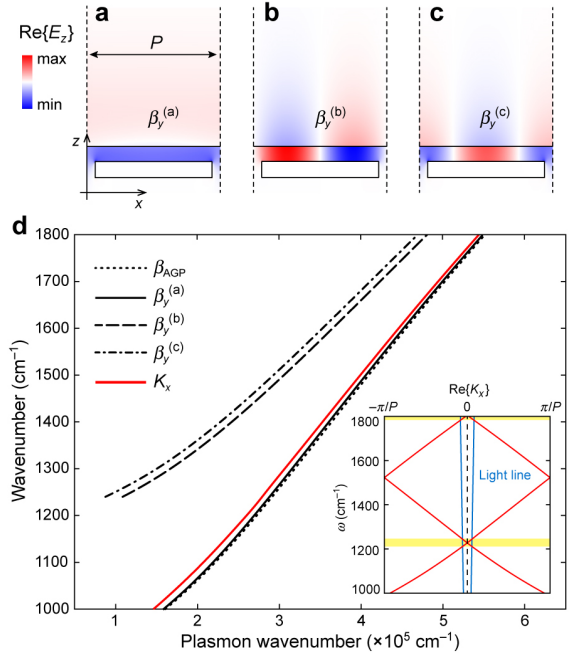

Figure 5. AGP dispersion in the nanoribbon array. **a-c** $E_z$ field distribution of AGP eigenmodes propagating along the nanoribbons in $y$-direction, obtained with the FEM eigenmode solver. **d** Numerically obtained dispersion of the AGP propagating along (black) and across (red) the nanoribbons axes; array parameters $P = 230$ nm and $E_F = 0.62$ eV correspond to the topical case analyzed in Fig. 6a,b. $\beta_{AGP}$ is the AGP wavenumber in the uniform structure, and $K_x$ is the Bloch wavenumber. Inset: band diagram of a planar AGP waveguide with an infinite 1D array of nanogaps, acting as partially reflective "mirrors"; yellow stripe indicates the free-space coupling.

The AGP dispersion in the $x$-direction is calculated using a simple model of a planar AGP waveguide with an infinite array of nanogaps, treated as partially reflective "mirrors" with complex transmission and reflection coefficients. Then, the dispersion solution is reduced to the eigenvalue problem for a lossy Bloch state in a 1D periodic medium[41] (see Supplementary Information section S-5). The real part of the numerically calculated Bloch wavenumber $K_x$ is shown in Fig. 5d (solid red line). The periodicity of the structure does not lead to the opening of a bandgap or flattening of bands at the center or edges of the Brillouin zone (inset in Fig. 5d), possibly due to the lossy nature of plasmonic modes. As a result, the density of optical states has similar value throughout the measured spectral range. Therefore, our measurements of the near-field contrast (Fig. 4) do not show any resonant feature at the frequency of the far-field array resonance corresponding to the phase matching. Since the wavevector of AGP is about two orders of magnitude larger than that of the free space, the radiative coupling to AGP can only happen near the Brillouin zone center, as indicated by the yellow stripe in the inset of Fig. 5d.

We proceed with analysis of the several instances of near-field images. The case of immediate interest is the phase matching between the array and the AGP. Figure 6a shows the spatial distribution of near-field signal intensity $s(x,y)$ and phase $\varphi(x,y)$ obtained at $\omega = 1200$ cm$^{-1}$ in the sample with $P = 230$ nm and $E_F = 0.62$ eV; $\lambda_{AGP} \approx 228$ nm, so that $\beta_{AGP} = 2\pi/P$. The measured and calculated profiles of $s \propto |E_z|$ and $\varphi \propto \arg\{E_z\}$ across the nanoribbons (Fig. 6b) both show a periodic variation across the nanoribbons with the period $P$ equal to the plasmonic wavelength, as demonstrated by the identical red scale bars on the left panel of Fig. 6a. Furthermore, the electric field amplitude (phase) has its maxima (minima) over the center of the nanoribbons, while the minima (maxima) are aligned with the nanogaps (Fig. 6b). Such field distribution corresponds to a resonant Bloch state (a standing wave) in a 1D array of partially reflective mirrors when $K_x = 2\pi/P$. Due to the phase matching in the array, the coupling to the AGP Bloch state is most efficient, thus it dominates the near-field signal.

When the AGP momentum starts to exceed that of the array, the near-field patterns of both amplitude and phase become completely different, as demonstrated in Fig. 6c for the sample with $P = 260$ nm, $\omega = 1150$ cm$^{-1}$, graphene $E_F = 0.52$ eV, and $\lambda_{AGP} \approx 225$ nm, so that $k_{AGP} = 1.15 \times 2\pi/P$. Now, the field maxima are recorded over the gaps, while the minima are at the center of the nanoribbons. We attribute such drastic change of the near-field distribution to the off-resonance excitation of multiple modes in the array, including the second-order array modes shown in Fig. 5b,c. As neither of the modes is dominant, they all contribute to the scattered near-field, which is particularly evident form the phase profile in Fig. 6d.

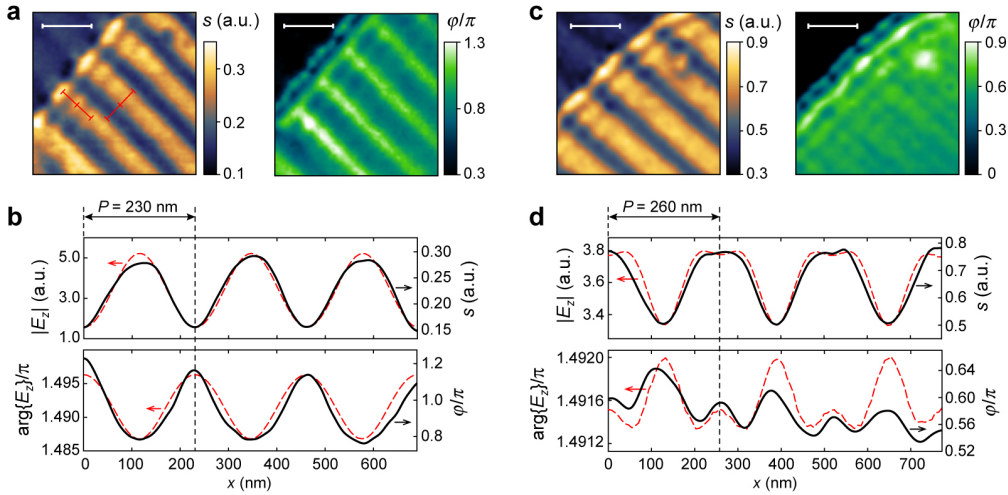

Figure 6. Near-field profile across the nanoribbons. **a** Near-field signal amplitude $s(x,y)$ and phase $\varphi(x,y)$ at the sample with $P = 230$ nm ($t = 18$ nm, $E_F = 0.62$ eV, $\lambda_{AGP} \approx 228$ nm) at $\omega = 1200$ cm$^{-1}$ when the AGP momentum $k_{AGP}$ is similar to the array momentum $2\pi/P$ (indicated by the identical red scale bars of 230 nm). **b** Profiles of $s(x,y)$ (top panel) and $\varphi(x,y)$ (bottom panel) shown in **a**, measured across the nanoribbons (black solid), and the numerically obtained by the full-wave simulation $|E_z|$ (top panel) and arg$\{E_z\}$ (bottom panel). **c** Same as in **a**, measured at the sample with $P = 260$ nm ($t = 18$ nm, $E_F = 0.52$ eV, $\lambda_{AGP} \approx 225$ nm) at $\omega = 1150$ cm$^{-1}$ when $k_{AGP} > 2\pi/P$. **d** Same as in **b**, showing the irregular near-field profile attributed to the mixed signal form the several array modes. White scale bars are 300 nm.

The near-field data in Fig. 6b,d is collected far from the graphene edge where the contribution of the edge-reflected AGP is minimized, which allows for employing a simple model for 2D full-wave simulations using an infinitely long line dipole instead of a point dipole. Yet the simulation results show a very good agreement with the measurements. At higher frequencies, when the AGP wavelength is significantly smaller than $P$ and $w$, the near-field mapping reveals the AGP reflection and scattering at nanoribbon edges and nanogaps. Even then, the simple numerical model predicts the near-field distributions with high accuracy (see Supplementary Information section S-6).

**DISCUSSION**

In conclusion, we employ the near-field coupling for the excitation and direct optical probing of AGP at mid-IR frequencies. Since the AGP field is tightly confined in the dielectric region underneath graphene, direct near-field optical imaging of the AGP fields has been considered very challenging. Nevertheless, with a highly sensitive s-SNOM system and a high-quality CVD graphene sample, we were able to directly probe the relatively weaker evanescent tail of AGP above the graphene layer. Furthermore, near-field imaging of mid-IR AGP reveals a very low damping rate even with unprotected CVD graphene at room temperature. The probed AGP mode is up to 2.3 times more confined than the GSP under similar conditions, yet exhibits a 1.4 times lower damping rate. These results highlight the importance of the AGP as a superior plasmonic species as compared to the GSP. Our investigation of the AGP in periodic nanostructures vividly illustrates the emerging AGP platform for strong light-mater interaction as a promising candidate for the future graphene-based optoelectronic devices operating in far- or near-field regimes.

**METHODS**

**Device fabrication**

Gold/alumina heterostructures on Si substrate were prepared with template-stripping method described in Ref.[6]. Large-area monocrystalline graphene was chemically grown on a single-crystal Cu foil. First, a commercial Cu foil (Nilaco Corporation, Japan) of 30 μm thickness was cut into ribbons and placed inside the CVD quartz tube, stretching between the hottest and the coldest zones inside the tube. Then, cycle annealing was introduced with a thermal gradient along the ribbons. The Cu foil was annealed at 1040 °C for two hours in an atmosphere of 40 sccm hydrogen and 1000 sccm argon gases. Then temperature was decreased to 700 °C during 30 min, and then increased up to 1040 °C during the same time. This process was repeated for four cycles in total, after which we opened the chamber to cool naturally.

For growing the high-quality graphene, we used the low-concentration methane (0.1% in argon) in four stages: ramping, annealing, growth, and cooling. First, the temperature was increased up to 1060 °C during one hour and then kept stable for one hour for annealing, which is necessary for removing organic molecules and enlarging the Cu grain size. Then, we used a mix of three gases ($CH_4$, $H_2$, Ar) for graphene synthesis. The graphene flake size is controlled by growth conditions such as the ratio between $CH_4$ and $H_2$ concentrations, the total amount of $CH_4$, and the growth time. Here, we purged 5 sccm of $CH_4$, 30 sccm of $H_2$, and 1000 sccm of Ar for a full coverage of Cu by graphene. Then, Cu ribbons were cut and graphene was wet-transferred from Cu onto the prepared heterostructures, and chemically doped by vapors of $HNO_3$ acid by placing the devices over the acid for 4 minutes.

**Device characterization**

The near-field scans were obtained by commercial s-SNOM (Neaspec GmbH) coupled with a tunable quantum cascade laser (Daylight Solutions, MIRcat), which illuminates the Pt-coated AFM tip (Nano World, ARROW-NCPt). The background-free interferometric signal[42], demodulated at third harmonic 3Ω (where Ω is the tapping frequency of the AFM tip), was used for near-field imaging. s-SNOM in AFM tapping mode was used to perform surface scans with 5 nm step.

**Numerical simulations**

Commercial FEM-based software (COMSOL Multiphysics) was used for full-wave simulations. In our 2D simulations, graphene is implemented as a thin film of finite thickness $\alpha = 0.2$ nm, having the effective relative dielectric permittivity $\varepsilon = \varepsilon_\mathrm{r} + i\sigma/(\omega\varepsilon_0\alpha)$, where $\varepsilon_\mathrm{r}$ is the background relative permittivity and $\sigma$ is the optical conductivity of graphene. Dielectric permittivity of gold was taken from Ref.[43], and that of thin film $Al_2O_3$ was taken from Ref.[44]. See Supplementary Information for details on full-wave simulations.

**ACKNOWLEDGEMENTS**

This work was supported by the Samsung Research Funding & Incubation Center of Samsung Electronics under Project Number SRFC-IT1702-14. S.G.M. acknowledges support from the Young Researchers program of the National Research Foundation of Korea (NRF) funded by the Korea government (MSIT) (2019R1C1C1011131). I.-H.L., T.L., and S.-H.O. acknowledge support from the U.S. National Science Foundation (NSF ECCS 1809723). S.-H.O. further acknowledges support from the Samsung Global Research Outreach (GRO) Program. S.L., T.-T.K., and Y.H.L. acknowledge support from the Institute for Basic Science of Korea (IBS-R011-D1). Device fabrication was conducted in the Minnesota Nano Center, which is supported by the U.S. National Science Foundation through the National Nano Coordinated Infrastructure Network (NNCI) under Award Number ECCS-2025124.

**AUTHOR CONTRIBUTIONS**

S.G.M, I.-H.L., and S.-H.O. conceived the idea. S.G.M. conducted the near-field measurements, analyzed the data, and wrote the manuscript. I.-H.L. fabricated the samples. S.L., T.-T.K., and Y.H.L. synthesized monocrystalline graphene. H.H. and J.T.H. assisted in samples preparation and measurements. T.L. and M.S.J. analyzed the data and wrote the manuscript. S.-H.O. and M.S.J. supervised the project.

**COMPETING INTERESTS**

The authors declare no competing interests.

Real-space imaging of acoustic plasmons in large-area CVD graphene

Sergey G. Menabde,[1,†] In-Ho Lee,[2,†] Sanghyub Lee,[3,4] Heonhak Ha,[1] Jacob T. Heiden,[1] Daehan Yoo,[2] Teun-Teun Kim,[3,5] Young Hee Lee,[3,4] Tony Low,[2] Sang-Hyun Oh,[2,*] and Min Seok Jang[1,**]

[1]School of Electrical Engineering, Korea Advanced Institute of Science and Technology (KAIST), Daejeon, Korea
[2]Department of Electrical and Computer Engineering, University of Minnesota, Minneapolis, USA
[3]Center for Integrated Nanostructure Physics (CINAP), Institute for Basic Science (IBS), Suwon, Korea
[4]Department of Energy Science, Sungkyunkwan University, Suwon, Korea
[5]Department of Physics, University of Ulsan, Ulsan, Korea
*sang@umn.edu
**jang.minseok@kaist.ac.kr
†These authors contributed equally to this work

## S-1. Numerical evaluation of AGP dispersion and damping rate

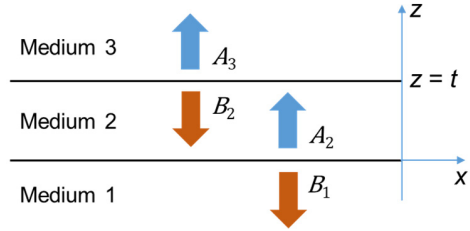

Figure S1. Schematics of a three-layer plasmonic structure. Arrows indicate the direction of exponential decay of the plasmonic mode magnetic field $H(z)$ having complex amplitudes $A_m$ and $B_m$.

The model of a uniform three-layer heterostructure with semi-infinite top and bottom layers is shown in Figure S1. The graphene layer is located at the interface between media "3" and "2" at $z = t$, where $t$ is the spacer thickness. Assuming the TM-polarized plasmonic mode propagating in the $x$-direction, the *ansatz* is made that the magnetic field in each medium $m = 1,2,3$ takes the form $\mathbf{H}^{(m)} = \hat{\mathbf{y}} H^{(m)}(z) e^{i[k_x x - \omega t]}$, where $k_x = 2\pi/\lambda_p$ is the wavevector component in the direction of propagation, and $\lambda_p$ is the plasmon wavelength. The magnetic field in each layer, exponentially decaying along the $z$-axis, can be presented as having constant complex amplitudes $A_m$ and $B_m$ in the following form:

$H^{(1)}(z) = A_1 e^{ik_z^{(1)} z} + B_1 e^{-ik_z^{(1)} z}, z < 0;$

$H^{(2)}(z) = A_2 e^{ik_z^{(2)}(z-t)} + B_2 e^{-ik_z^{(2)}(z-t)}, 0 < z < t;$

$H^{(3)}(z) = A_3 e^{ik_z^{(3)}(z-t)} + B_3 e^{-ik_z^{(3)}(z-t)}, z > t.$

In the above equations, $k_z^{(m)} = \sqrt{k_0^2 \varepsilon_m - k_x^2}$ is the $z$-component of the plasmon wavevector in medium $m$, $k_0 = \omega/c$ is the free-space wavevector, and $\varepsilon_m$ is the dielectric permittivity. Then, substituting the equations for the magnetic field into Maxwell's curl equation $\nabla \times \mathbf{H} - \varepsilon \varepsilon_0 \, \partial \mathbf{E}/\partial t = \mathbf{J}$, we apply field continuity conditions for electric and magnetic fields at each interface: $E_x^{(m+1)} - E_x^{(m)} = 0$ and $H^{(m+1)} - H^{(m)} = \mathrm{K}$, where $K(z=0) = 0$ and $K(z=t) = \sigma(\omega)E_x$; $\sigma(\omega)$ is the optical conductivity of graphene. Taking $A_1 = B_3 = 0$ (absence of reflected or incident waves in media "1" and "3"), we obtain a system of four linear equations for the four unknowns $A_m$ and $B_m$. By searching for all possible solutions of the system of equations (if any), all plasmonic eigenmodes supported by the system can be found. This powerful approach can be used to evaluate eigenmodes of multilayer heterostructures with an arbitrary number of layers, or to calculate complex reflection coefficients in the presence of incident wave (e.g. when $B_3 = 1$).

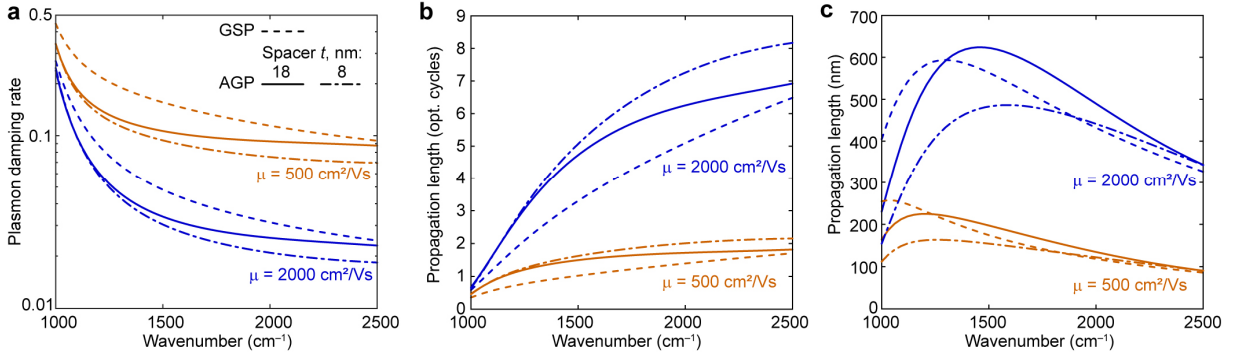

Figure S2. **a** Plasmon damping rate $\text{Im}\{q\}/\text{Re}\{q\}$, where $q$ is the plasmon propagation constant. **b** Plasmon propagation length in optical cycles. **c** Plasmon propagation length in nm. Analytically calculated results for: AGP mode with spacer thickness 18 nm (solid; $E_F = 0.51$ eV) and 8 nm (dash-dot; $E_F = 0.49$ eV), and GSP mode on a semi-infinite alumina (dashed); for the carrier mobility in graphene 2000 cm²/Vs (blue) and 500 cm²/Vs (orange).

Figure S2a demonstrates the damping rates calculated for the AGP eigenmode in samples with 18 nm-thick (solid) and 8 nm-thick (dash-dot) alumina spacer with the graphene carrier mobility (2000 cm²/Vs; blue) and Fermi level as obtained from the s-SNOM measurements at 1150 cm⁻¹ (Figure 2b,c in the main text). For comparison, the calculated damping rate for GSP on a semi-infinite alumina substrate is also shown (solid), as well as the damping rates for all modes for when the carrier mobility is 500 cm²/Vs (orange). AGP dispersion corresponds to the sample. Figure S2b shows the propagation length of the AGP and GSP modes shown in Figure S2a, calculated in optical cycles as $l = \text{Re}\{q\}/(2\pi\text{Im}\{q\})$; and Figure S2c shows the same propagation length in nm. From Figure S2a, it is evident that the AGP is always less damped than the GSP in the mid-IR, irrespective of the graphene doping level or spacer thickness. Even more surprisingly, the AGP becomes less damped when the spacer thickness decreases, despite larger mode confinement factor (as discussed in the main text). This effect is more pronounced at higher frequencies where the dispersion of AGP in the thinner spacer deviates from the GSP mode, while the less confined AGP in the 18 nm-thick spacer starts to merge into the GSP as the influence of the gold layer on diminishes.

**S-2. Impact of surface quality on AGP scattering**

Figure S3 shows the atomic force microscope (AFM) scans of surfaces in the sample fabricated by the template stripping method: gold surface under the alumina spacer (that was removed from the pre-fabricated sample), and the top surface of alumina on top of which graphene was deposited. Naturally formed grain structure of thermally evaporated gold film have an average grain size of 80 nm. The same grain size is measured at the alumina surface, as the ALD deposition of $Al_2O_3$ is conformal. The RMS surface roughness $a_{RMS}$ measured across a 1×1 μm² area is approximately 0.5 nm in both cases, which is very close to a typical roughness of a Si wafer 0.3 nm.

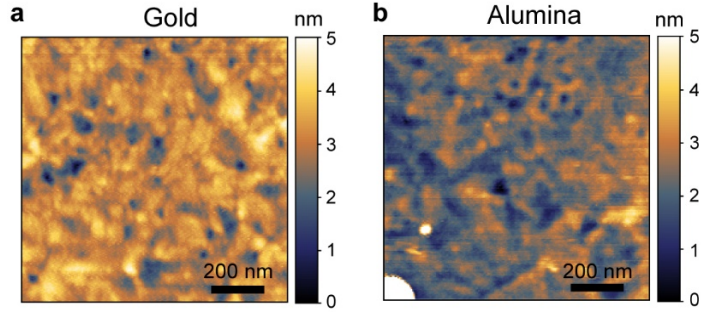

Figure S3. AFM scans of the gold and alumina surfaces between which the propagating AGP mode is confined. The sample was fabricated by template stripping.

In order to estimate the AGP and GSP loss caused by scattering on rough surfaces, we performed numerical finite-element method (FEM) simulations of propagating plasmons in a uniform structure where surfaces are modelled as having a wave-like (sinusoidal) profile, with the period twice larger than the average grain size, and peak-to-peak distance given by $2\sqrt{2}a_{RMS}$. Figure S4 demonstrates the analytically calculated plasmon propagation length (solid lines) measured in optical cycles, $l = \text{Re}\{q\}/(2\pi\text{Im}\{q\})$, as a function of frequency for parameters corresponding to the case shown in Fig. 2b,c in the main text, and neglecting the scattering loss. Also shown are the numerically obtained $l$ in the structure with smooth (black circles), and corrugated interfaces with $a_{RMS}$ = 0.5 nm (red diamonds) and $a_{RMS}$ = 2 nm (blue crosses). We find no significant change of the AGP propagation length even when the surface roughness is 2 nm, particularly at the excitation frequencies in our experiments 1000-1300 cm$^{-1}$. Therefore, we conclude that the scattering loss of AGP in our samples can be neglected when analyzing the near-field data.

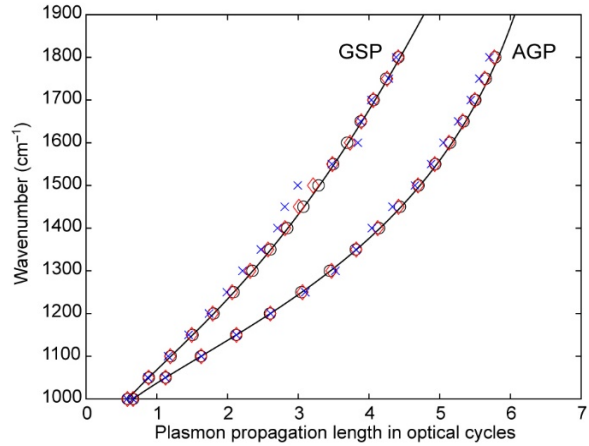

Figure S4. Plasmon propagation length in units of optical cycles. Analytically (solid lines) and numerically obtained data (black circles) are plotted together with numerical data for $a_{RMS}$ = 0.5 nm (red diamonds) and for $a_{RMS}$ = 2 nm (blue crosses). Heterostructure model parameters correspond to the case analysed in Figure 2b,c in the main text.

**S-3. Numerical simulation of AGP fringes at graphene edge**

As explained in the main text, the AFM tip efficiently scatters only the *z*-component of the electric field ($E_z$) due to its vertically elongated geometry. Therefore, the intensity of the scattered near-field signal is directly proportional to the $E_z$ above the sample at the position of the tip. The interaction between the point dipole source (i.e., the AFM tip of the s-SNOM) and the sample (i.e., the heterostructure supporting AGP) can be numerically modelled by FEM or finite difference time domain (FDTD) methods in a quasi-static approximation, since the mechanical oscillation frequency of the AFM tip (~100 kHz) is several orders of magnitude lower than the optical frequencies (~10 THz). Hence, by calculating a steady-state solution for the electromagnetic fields at different tip positions, it is possible to reconstruct the near-field signal scattered by the AFM tip, $s \propto |E_z|$, where $E_z$ is the vertical component of the electric field under the tip. In the presence on the graphene edge, probed $E_z$ is a result of superposition between the locally induced field (due to the interaction between the tip and the heterostructure) and the field of the propagating AGP mode that has been reflected from the graphene edge.

Figure S5 shows the schematics of the model used to calculate the AGP fields. The actual height $h_1$ of the induced electric dipole over the sample is unknown, as well as the height $h_2$ of the tip area effectively scattering the AGP fields. Another unknown parameter is the reflection coefficient $R$ of the graphene termination, which is ≈0.99 in a structure with an ideal edge. These three parameters, along with the optical conductivity of graphene, can be determined by fitting the simulations data because each combination of parameters produces a unique near-field intensity pattern. Our calculations have provided the following model parameters corresponding to the near-field data shown in Figure 2b in the main text: $h_1$ = 160 nm, $h_2$ = 85 nm, and $R$ = 0.8. Since the simulations are done in 2D for the sake of high speed and low resource consumption, the electric field of the reflected AGP must be adjusted to consider the energy conservation in a diverging circular wave: $E_{AGP} \propto r^{-0.5}$, where $r$ is the propagation distance from the excitation point.

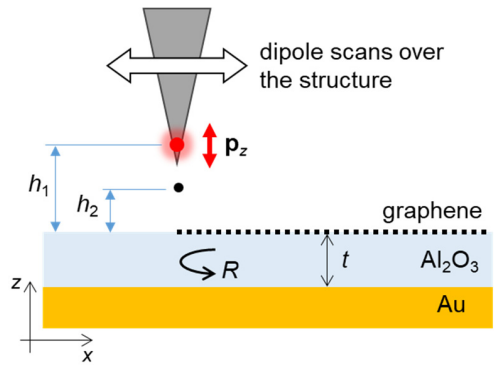

Figure S5. Numerical simulation model to calculate the near-field signal intensity detected by s-SNOM. AFM tip is simulated as a *z*-oriented point dipole source, and scattered field intensity is proportional to $|E_z|$ at some point under the dipole. Propagating AGP experiences reflection at the graphene edge with reflection coefficient $R$.

**S-4. s-SNOM-measured AGP dispersion at samples with gold nanoribbons**

In order to determine the Fermi level in the samples with patterned gold, we use dispersion measurements taken over the area of interest. Figure S6 shows the AGP dispersion obtained from the plasmon fringes at near-field images (circles) and fitted by analytical dispersion (i.e. by the value of $E_F$ in graphene) for graphene placed on the alumina spacer over gold nanoribbons of different width (as denoted). All data were gathered from the samples with an 18 nm-thick spacer. Due to the use of chemical doping, the Fermi level of graphene has to be evaluated for every set of measurements taken at different sample area and time. The obtained value of the Fermi level was used to analyze the near-field images shown in the main text.

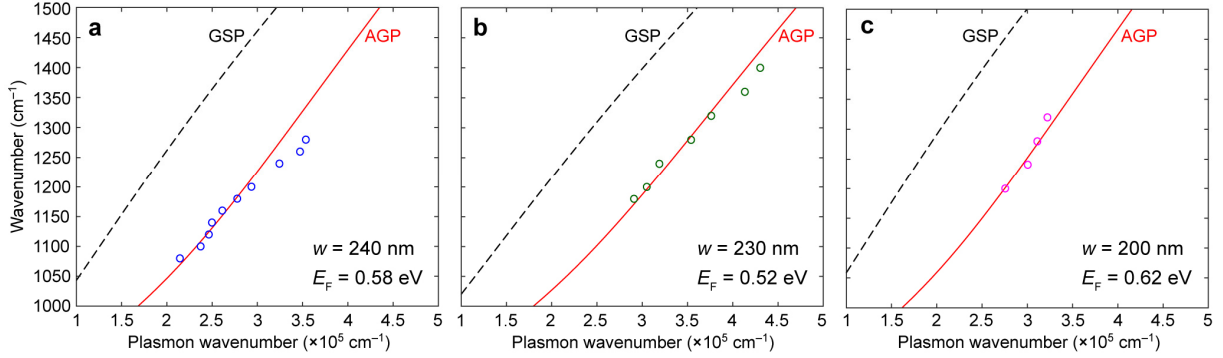

Figure S6. Analytical AGP dispersion (solid red) calculated for the fitted value of graphene Fermi level (as noted). Analytical dispersion is fitted based on the measured AGP wavelength (circles) for samples with different width of gold nanoribbons (as noted). Also shown, GSP dispersion (dashed black) for graphene with a given Fermi level if placed on a semi-infinite alumina substrate.

**S-5. AGP dispersion for the propagation across the ribbons**

To calculate the dispersion for AGP mode propagating across the ribbons (along the *x*-axis), we employ a simple model shown in Fig. S7a. In this model, the nanogap is treated as a partially reflective lateral mirror for the propagating AGP mode with complex reflection and transmission coefficients $\alpha(\omega)$ and $\delta(\omega)$, respectively. Then, the dispersion solution reduces to the eigenvalue problem for a 1D photonic crystal with the lattice constant $P$: $\det[\mathbf{M}_0 - e^{iKP}\mathbf{I}] = 0$, where $K$ is the Bloch wavenumber and $\mathbf{M}_0$ is the wave-transfer matrix for the unit cell. For the model presented in Fig. S7a, the wave-transfer matrix is given by:

$$\mathbf{M}_0 = \frac{1}{\delta}\begin{bmatrix} e^{i\varphi}(\delta^2 - \alpha^2) & \alpha e^{i\varphi} \\ -\alpha e^{-i\varphi} & e^{-i\varphi} \end{bmatrix},$$

where $\varphi = wk_{AGP}$ is the accumulated AGP phase over the nanoribbon width. Reflection and transmission coefficients are numerically obtained from the full-wave FEM simulations (COMSOL), where AGP is launched at a numerical port boundary and propagates over a single nanogap in the gold layer. Amplitude and phase of the reflection and transmission coefficients for lossy (realistic) and low-loss cases are shown in Fig. S7b. Note that the amplitude of transmission is much larger than the amplitude of reflection. Substituting the wave-transfer matrix into the eigenvalue problem, we obtain the following general solution for the Bloch state:

$$2\cos(KP) = \frac{e^{i\varphi}}{\delta}(\delta^2 - \alpha^2 + e^{-2i\varphi}).$$

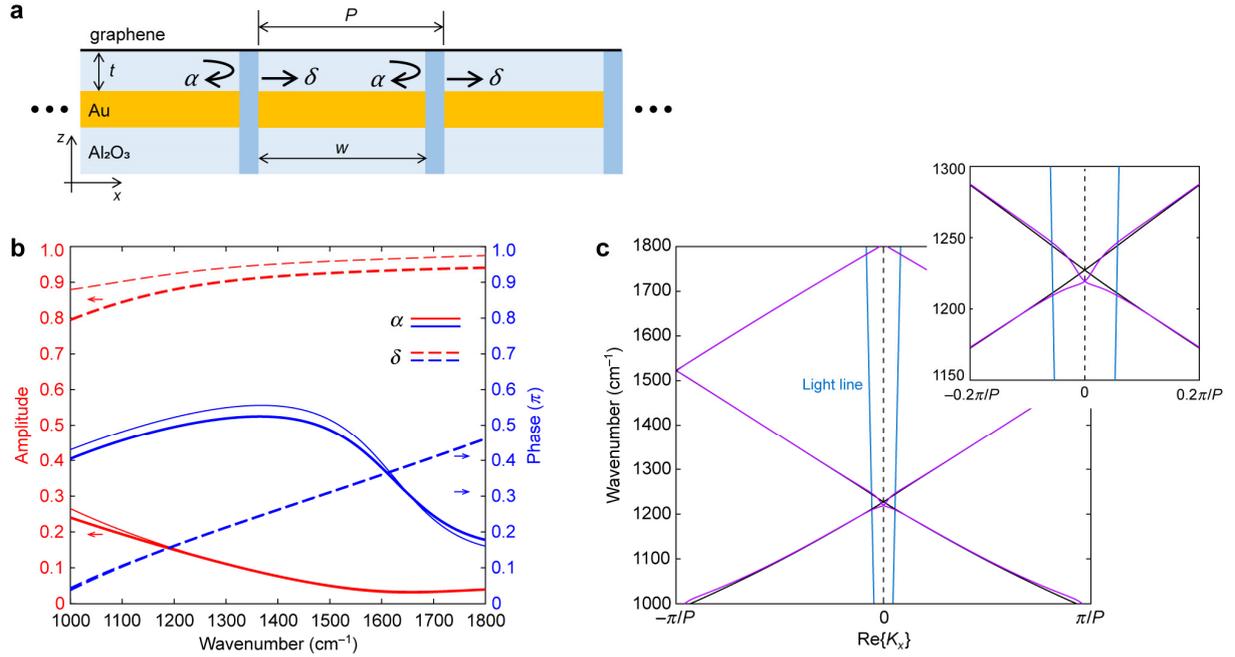

Figure S7. **a** Model of 1D photonic crystal configuration for AGP where the nanogap between gold bars is treated as an element with complex reflection and transmission coefficients $\alpha(\omega)$ and $\delta(\omega)$, respectively. **b** Numerically calculated $\alpha(\omega)$ (solid) and $\delta(\omega)$ (dashed) of the 30 nm-wide nanogap in a 20 nm-thick gold layer: amplitude (red) and phase (blue). Thick lines correspond to real materials, while thin lines correspond to the low-loss materials. **c** AGP band diagram for the 1D array of nanogaps for the actual materials (black; shown in the main text) and the low-loss materials (purple); close-up shows the center of the Brillouin zone where bands start to curve.

As discussed in the main text and illustrated by Fig. 5, the band structure for the 1D periodic array of gaps does not show any bandgap. We also calculated the structure with tuned-down losses in all materials: with the carrier mobility in graphene 20,000 cm$^2$/Vs, purely real alumina permittivity, and gold permittivity calculated according to the Drude-Sommerfeld model (Ref.[44]) with electron relaxation time 100 times larger than the reported experimental value of 14 fs. The 1D array band structures for the actual and low-loss materials are shown in Fig. S7c, and demonstrates the bands bending at the centre of the Brillouin zone.

### S-6. Near-field profile over nanoribbons array at high frequencies

Figure S8 demonstrates the distribution of near-field amplitude $s$ and phase $\varphi$ over the sample with $P = 260$ nm ($w = 230$ nm) shown in Figure 6a in the main text, but at higher excitation frequencies of 1240 cm$^{-1}$ (Fig. S8a) and 1400 cm$^{-1}$ (Fig. S8c). At these frequencies, the scattering of the AGP at the gold ribbon edges intensifies, and may contribute to new features in the near-field profiles across the ribbons (solid black in Fig. S8b,d). These features are not observed at the simulated data plots (dashed red in Fig. S8b,d), obtained by the 2D full-wale simulations with a source being a line of dipoles, without adjusting the AGP electric field amplitude for a circular diverging wavefront as has been done in the simulation of the interference fringes at the graphene edge in section S-3. Note that the position of the field maxima/minima is different in each case.

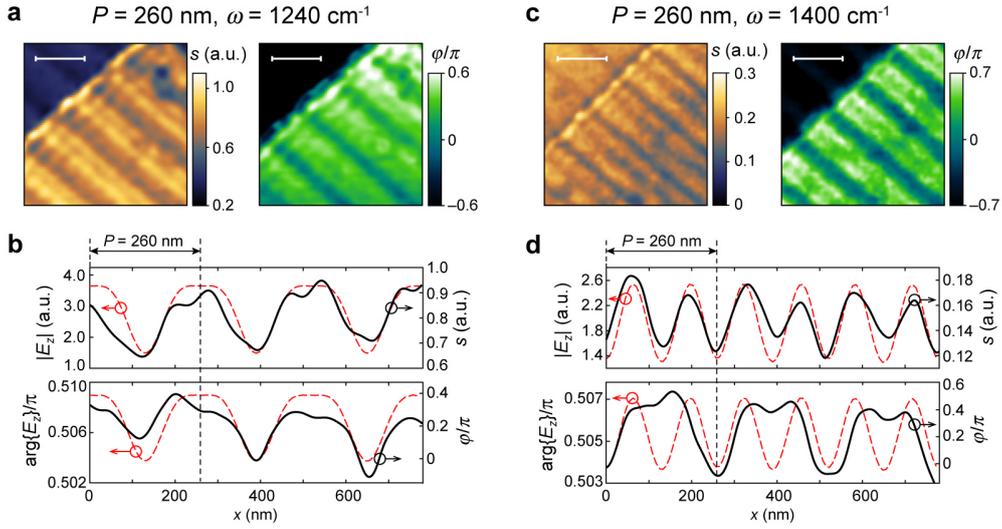

Figure S8. **a** Near-field signal amplitude $s(x,y)$ and phase $\varphi(x,y)$ measured over the sample with $P = 260$ nm ($t = 18$ nm, $E_F = 0.52$ eV, $\lambda_{AGP} \approx 225$ nm) at $\omega = 1240$ cm$^{-1}$ when the AGP momentum $k_{AGP}$ is larger than the array momentum $2\pi/P$. **b** Profiles of $s(x,y)$ (top panel) and $\varphi(x,y)$ (bottom panel) across the nanoribbons (black solid) corresponding to the scans shown in **a**, and the numerically obtained $|E_z|$ (top panel) and $\arg\{E_z\}$ (bottom panel) from the FEM simulation of the near-field measurement; parameters of the tip model are the same as in Figure 2c. **c,d** Same as in a,b at $\omega = 1400$ cm$^{-1}$ when the AGP momentum is significantly larger than the array momentum. Scale bars are 300 nm.

20